\documentstyle[aps,epsf]{revtex}
\begin{document}
\title{Topological interactions in DNA catenanes}
\author{M.Otto and T.A.Vilgis}
\address{Max-Planck-Institut f\"ur Polymerforschung\\ 
Postfach 3148, D-55021 Mainz, Germany}
%
%
\maketitle
\begin{abstract}

The elasticity of DNA catenanes, i.e. multiply linked DNA rings, is 
investigated using the Gauss invariant as a minimal model for 
topology conservation. An effective elastic free energy as a function 
of the distance $R$ between segments located on different rings is obtained. 
An anharmonic part at large distances, 
growing as $R^{4}$, if $R\gg R_{\rm G}$ 
($R_{\rm G}$ being the radius of gyration of a random walk ring) 
is found, while for $R\ll R_{\rm G}$ the interaction 
is strongly repulsive. Treating the attractive interaction as the dominant
one, distribution functions for the distance between segments located on 
different rings for several linking numbers are derived which are in 
qualitative agreement with distributions functions obtained 
experimentally from 
electron micrographs of DNA catenanes (S. D. Levene et al., {\it Biophys.J.} {\bf 69}, 277, 1995).

\end{abstract}
%
%
%
%
Knots and links made from DNA beautifully elucidate the general
role of topology in nature. 
More specifically,
DNA rings from bacteria such as Escherichia coli form so-called 
catenanes (the chemical term for entangled rings called links by 
knot theorists) as intermediate products of
DNA replication and recombination \cite{wasserman:86}. 
Together with single knotted DNA rings they
form the class of so-called topoisomers, i.e. topologically distinct isomers. 
They can be isolated experimentally 
by manipulating certain enzymes (topoisomerases) switching back between 
various topologies \cite{wasserman:86,seeman:93}. 
Several catenanes  and knots have been
identified e.g. by electron microscopy and electrophoresis 
\cite{wasserman:86,adams:92}. Most recently, the conformation of 
open-circular, multiply linked dimeric DNA catenanes has been studied by 
electron microscopy \cite{levene:95}. From the tracing of the molecular 
contours on the micrographs, in particular, 
distribution functions for the distance (in the projection plane) between
segments on separate rings have been established. As the linking number 
- measuring the degree of concatenation between pairs of rings - 
is increased, 
the catenane conformation becomes more and more compact due to an 
effective attractive interaction - of topological origin - between segments.

In this letter, we present a theory to explain the experimental results of 
ref.\cite{levene:95}
mentioned above.
From a theoretical point of view, the statistical mechanics 
of entangled rings 
is a generally unsolved problem \cite{edwards:67,edwards:68}. 
The essential difficulty is how to specify the topological state 
of the system, which is assumed to remain unchanged 
with respect to conformational fluctuations. The mathematical 
answer to this question is 
a so-called link invariant, various types of which are discussed in 
knot theory while none of them is one-to-one 
\cite{kauffman:93,adams:94}. The most simple, yet most crude invariant 
is the Gauss integral for two given closed loops $\alpha$
and $\beta$
\begin{equation}
\label{1.1}
\Phi(C_\alpha, C_\beta)\equiv \Phi_{\alpha\beta}
=\frac{1}{4\pi}\oint_{C_\alpha}ds\oint_{C_\beta}ds'
\dot{{\bf r}}^\alpha (s)\wedge\dot{{\bf r}}^\beta (s')\cdot
\frac{{\bf r}^\alpha (s)-{\bf r}^\beta (s')}
{|{\bf r}^\alpha (s)-{\bf r}^\beta (s')|^3}, 
\end{equation}
which is a double line integral over the loop contours $C_{\alpha}$, $C_{\beta}$
where ${\bf r}^\alpha (s)$ denotes the segment position in 
$d=3$ space dimensions as a function of the
chain contour. Unfortunately, the Gauss invariant has 
the same value for two unlinked rings
and the so-called Whitehead link, and is therefore not one-to-one. In fact 
a whole series of higher-order linking coefficients (HOLC) needs to 
considered in order to uniquely characterize the link topology.
A different type of invariants, known as knot (or link) 
polynomials and defined via
skein relations, are suitable for computer simulations 
\cite{volod:74,volod:75,koniaris:91,deguchi:93:1,deguchi:93:2}, but not 
for an analytical theory as they are 
independent of the segment positions and therefore cannot be coupled to 
a particular polymer model. Recently however, a deep connection between the 
Gauss invariant as well as HOLC
and polynomials for knots and links has been established 
\cite{witten:89}. As a consequence, it has been demonstrated 
that the Gauss invariant is a first order
approximant to the true topology of links \cite{guada:93}. It is
however no invariant for knots \cite{calu:59,volod:74} as was supposed in 
\cite{edwards:68,deam:76}. 

In view of the experimental data, however, we have
additional reasons to use the Gauss invariant more safely. The dimeric DNA
catenanes are only interlinked and the individual rigs are not self linked.
Moreover, the Gauss invariant has proven numerically to be 
very efficient for simple models describing the polymer conformation, 
such as the 
random walk and the self-avoiding walk 
\cite{everaers:95,everaers:96,auhl:95}. It was especially shown
that almost 95\% of the conformations could be
found by using the Gauss invariant, compared to the
resulting number obtained by the simulation \cite{auhl:95}. 

Given these facts about the Gauss invariant, we propose to use it as a minimal 
model for topology conservation in polymer ring systems. 
In what follows we develop a framework for an arbitrary number of concatenated
rings. 
Finally, 
in order to make contact with the experiments on DNA catenanes, we restrict 
our analysis to two concatenated rings. We will consider 
random walk rings with {\it no} additional 
local 
excluded volume interactions other than those imposed 
by topology conservation. This procedure may appear strange, but it assures
the rigorous applicability of the Gauss invariant in order to classify the 
pair-wise topology of rings as self-intersections are allowed for a single
ring. This ansatz neglects excluded volume interactions of a ring with itself,
but maintains effectively
the excluded volume interactions between different rings 
as a result of
topology conservation. The latter is imposed by fixing the linking number 
matrix
$m_{\alpha\beta}=\Phi_{\alpha\beta}$ between rings $\alpha$ and $\beta$.
Furthermore, we keep one point fixed on each ring, i.e. ${\bf r}^{\alpha}_{0}$ for 
every ring $\alpha$. Our goal is then to calculate the partition function 
$Z(m_{\alpha\beta};\{{\bf r}^{\alpha}_{0}\})=\langle 
\delta(m_{\alpha\beta}-\Phi_{\alpha\beta})
\rangle_{\{{\bf r}^{\alpha}_{0}\}}$, taking the average over random walk ring
conformations with the given constraints.
The use of a random walk as a conformational model for DNA rings is of course
insufficient as the local bending stiffness is neglected. It has been shown
however, that for large chain lengths, DNA rings without supercoils behave
like Gaussian rings \cite{marko:95} which corresponds to the experiment in 
\cite{levene:95} discussed at the end of this letter.

We now present the general route to calculate
$Z(m_{\alpha\beta};\{{\bf r}^{\alpha}_{0}\})$ (the details will be presented 
elsewhere). The two ring partition function has to be computed subject to the
constraint of the topology conservation. The latter is expressed by a simple
delta function constraint, which maintains the 
linking number $m_{\alpha \beta}$ between the two
rings, using the Gaussian invariant. A Hubbard - Stratonovich transformation
handles the Gaussian integral and introduces a conjugated field theory.  
The partition function is then Fourier transformed to
$Z(s_{\alpha\beta};\{{\bf r}^{\alpha}_{0}\})$ where $s_{\alpha\beta}$ is the
symmetric 
part of the matrix conjugate to $m_{\alpha\beta}$ (the general 
approach to define the problem for many-rings systems is out-lined in detail 
and solved for $m_{\alpha\beta}=0$
in ref.
\cite{brereton:95}). In the conjugate partition function
$Z(s_{\alpha\beta};\{{\bf r}^{\alpha}_{0}\})$, the Gauss invariant appears as an 
effective (imaginary) interaction which can be re-expressed using collective 
variables, i.e. the tangent vector density of ring $\alpha$, 
${\bf j}^{\alpha}({\bf x})=\oint ds \dot{{\bf r}}^{\alpha}(s)\delta({\bf x}-{\bf r}^{\alpha}(s))$.

We then use the fact that a symmetric matrix 
$s_{\alpha\beta}$ can be decomposed 
according to
$s_{\alpha\beta}=\sum_{\gamma}\eta_{\alpha\gamma}\eta_{\gamma\beta}$, and 
define a modified tangent vector density according to
$\Psi^{\alpha}=\sum_{\beta}\eta_{\alpha\beta}{\bf j}^{\beta}$.  
The variables $\Psi^{\alpha}({\bf x})$ are introduced into 
the partition function $Z(s_{\alpha\beta};\{{\bf r}^{\alpha}_{0}\})$ 
using conjugate gauge fields
${\bf B}^{\alpha}({\bf x})$. The modified tangent vector densities $\Psi^{\alpha}$ are 
finally summed over to give the following conjugate partition function:
\begin{eqnarray}
\label{def1}
Z(s_{\alpha\beta}; \{{\bf r}^{\alpha}_0 \} )&=&
\int \prod_{\alpha}\prod_{\bf k\neq 0}d{\bf B}^{\alpha}({\bf k})
\exp\left(-\sum_\alpha
\int_{\bf q} B^\alpha_\mu({\bf q}) B^\alpha_\nu({\bf -q})
\epsilon_{\mu\lambda\nu}q_\lambda\right)\times\nonumber\\
&&\left\langle
\exp\left(
-i\sum_\alpha\int_{\bf q}\hat{{\bf \Psi}}^\alpha({\bf q})
\cdot{\bf B}^\alpha({\bf -q})
\right)
\right\rangle_{\{{\bf r}^{\alpha}_0 \}}
\end{eqnarray}
Here $\epsilon_{\mu \lambda \nu}$ is the standard Levi - Cevita symbol.
Thus, the linking number constraint has been rewritten in terms of an
$n_{p}$-component Abelian Chern-Simons field theory. In the case of 
DNA catenanes, $n_{p}=2$. The next step is to perform the average over 
the random walk rings $\{{\bf r}^{\alpha}(s)\}$ while keeping fixed the points 
${\bf r}^{\alpha}_{0}$ on each ring $\alpha$. To simplify the functional
integration, a preaveraging procedure is carried
out for Fourier transformed tangent vector 
densities ${\bf j}^{\alpha}({\bf q})$:
\begin{eqnarray}
\label{def4}
j_\mu^\alpha({\bf q})
&\simeq&\oint ds \dot{r}^\alpha_\mu(s) 
e^{-i{\bf q}\cdot{\bf r}^\alpha_0}
\langle
e^{-
i{\bf q}\cdot({\bf r}^\alpha(s)-{\bf r}^\alpha_0)}\rangle\nonumber\\
&=&\oint ds \dot{r}^\alpha_\mu(s) 
e^{-i{\bf q}\cdot{\bf r}^\alpha_0}
e^{-\frac{l^2}{2d}q^2s}
\end{eqnarray}
The preaveraging approximation is not critical in the physical problem
discussed here. The topology is conserved and the main physical contribution
is taken already by the average.
The integration over the gauge fields in Eq.(\ref{def1}) is approximated by
 a cumulant expansion (as a power series in $s_{\alpha\beta}$), keeping terms 
up to $O(s_{\alpha\beta}^{2})$. Finally, we obtain the partition function for 
fixed linking number $Z(m_{\alpha\beta};\{{\bf r}^{\alpha}_{0}\})$ (after 
integrating over $s_{\alpha\beta}$).

The resulting effective free energy for two interlinked rings 
$\alpha$ and $\beta$ (to order $O(m^2_{\alpha\beta})$ in the linking 
number) reads:
\begin{equation}
\label{def16}
\beta F_{\alpha\beta}(R=|{\bf r}^\alpha_0-{\bf r}^\beta_0|)=
\frac{m^2_{\alpha\beta}}{4}
(2\pi)^{3}\left(\sum_{j=1}^\infty
\frac{(-1)^j}{(j-1)!(2j+1)}\left(\frac{R}{2\sqrt{2}R_G}\right)^{2j-1}
\right)^{-2}
\end{equation}
The sum can be evaluated exactly by re-summation and transformation to an
integral over a exponential function. This yields an expression which contains
an Error function. To compare our result to the experimental situation,
seen in electron micrographs we need the to project the three dimensional
result of the corresponding two dimensional case. Therefore we take the
asymptotics of the free energy for large and small values of $R$
Introducing the scaling variable $x=R/R_{\rm G}$, Eq.(\ref{def16}) 
is approximated by a simple asymptotic form (see also FIG.1). 
\begin{equation}
\label{def19}
\beta F_{\alpha\beta}(x)=
\frac{m^2_{\alpha\beta}}{4}\left((2\pi)^3 \frac{32}{x^2}+2\pi^2x^4\right)
\end{equation}
The second term in Eq.(\ref{def19}) is asymptotically exact for $x\gg 1$ and 
produces an anharmonic term  on distances $R\gg R_{\rm G}$. The result can
be intuitively understood. The two linked rings must -  upon relative 
deformation with respect to each other produce a stronger 
elasticity than a Gaussian chain.

To understand the behavior of the topological free energy let us come back to
simple tube arguments for entangled polymers.  Although the rings are confined
strongly by entanglements the elasticity is purely governed by entropy. Thus
let us estimate the number of configurations of the two rings when $m$ is
large and the two rings are linked many times. In terms of a tube like model
the distance between two links form a primitive path step \cite{doi,edvil}.
The mean square primitive path step length is simply given by the mean
distance between two entanglements, i.e., $a^{2} = l^{2}N/m$. The total number
of constraints is the linking number $m$ themselves. Both yield a modulus that
is proportional to the square of the linking number, which is reflected by the
free energy calculated above. To understand the peculiar $x^{4}$ behavior let
us estimate the linking constraint for the two rings, eq.(\ref{1.1}). The
lowest order cumulant of relevance is revealed by $\langle \Phi^{2}
(C_{\alpha}, C_{\beta})\rangle$, where the average is taken by the Gaussian
configuration. By estimating crudely the bond vectors as unit vectors and
taking into account the cross product, the average is of the order of
$N^{2}/R^{4}$. The linking number constraint must be incooperated $\langle
\delta(m-\Phi(C_{\alpha},C_{\beta}))\rangle$ to the partition function.
Parameterizing the delta function by an auxillary integral, and carring out
the lowest order average yields the inverse of the estimate of the second
cumulant, i.e., $\exp(- m^{2}R^{4}/l^{4}N^{^{2}})$, whis is calculated above.
Moreover, the result $F \propto x^4$ to this order agrees nicely with the
estimation of the limits to such exponents presented by Frisch and Wasserman
\cite{wasserman2} if two linked rings are pulled against each other. Knot
theory \cite{rolfson} suggest limits for the exponent between 2 and 4.  Higher
order approximations yield higher power terms, but it they can be related to
the "finite extensibility" of the entangled object \cite{edvil} and are of
little  importance here.

The
first term in eq.(\ref{def19})
gives the repulsive interaction on small scales due to the
increased probability that a segment ${\bf r}^{\alpha}_{0}$ of one ring meets
segments of the other one in a neighborhood close to the link.  Its explicit
form behaving as $1/x^{2}$ is due to an approximated evaluation of an integral
over ${\bf k}$ space keeping only the first-order approximant.  Intuitively
the appearance of such a term is also not surprising. It corresponds to the
confinement energy of Gaussian structures. At smaller values of $x$ topology
plays of the dominant role.  Thus we expect the corresponding confnement term
$F_{\rm conf} \propto (R_{\rm G}/R)^{2}$ for the compression of the two
entangled rings.  As higher order terms are considered, the repulsive tail of
the free energy Eq.(\ref{def19}) approaches the vertical axis with presumably
a delta peak being the limiting value while the $x^{4}$-term persists.
Therefore we suppose that the latter term contains the essential physics of
pairs of linked rings whereas the repulsive interaction is only relevant in
the case that both rings are strongly squeezed together.

We now proceed to calculate the distribution function for distances between 
segments located on different rings that form a DNA catenane. 
As suggested by Levene et al. \cite{levene:95}, we assume that the electron 
micrographs of DNA catenanes display a 2-dimensional projection of the real 
3D conformation. We therefore start from the following three-dimensional 
distribution function
\begin{equation}
\label{dna1}
P(x)\sim {\cal N}\exp \left(-m^2 x^4\right),
\end{equation}
the variable $x$ being a scaled distance (absorbing also the numerical 
factors) between two segments located on 
different rings, and $m$ the linking number. Then, cylindrical coordinates 
${\bf r}=(z,r,\varphi)$ are used with $x^{2}=z^{2}+r^{2}$. Finally, 
the normalized 2D 
distribution function is obtained:
\begin{equation}
\label{dna5}
P(r)=Cm^{3/2}r\exp\left(-\frac{m^2}{2}r^4\right)D_{-1/2}\left(
\sqrt{2}mr^2
\right)
\end{equation}
The function $D_{-1/2}(u)$ is a parabolic-cylinder function 
\cite{gradstein:81}. The constant $C$ is a numerical constant subject to 
the constraint that $\int_{0}^{\infty}dr P(r)=1$. A selection of 2D 
distribution functions for linking numbers $m=3, \;5,\; 7,\; 9$ is given
in FIG. 2. They corresponding curves
show a remarkably good qualitative agreement with the 
respective distributions that are determined experimentally in
\cite{levene:95}. Their result is reproduced in FIG. 3. 
The essential characteristics of this family of 
functions are that for increasing
linking numbers, their peak values occur for lower and lower values of $r$. 
Moreover, the distribution functions become more and more narrow as $m$
increases. This clearly indicates that catenanes of higher linking numbers 
show a more compact conformation. 
Concerning a more quantitative comparison between our theoretical 
work and the experimental results obtained in \cite{levene:95}, the
semiflexible nature of DNA catenanes must be taken into account
\cite{marko:94,marko:95:1,marko:95:2}. The appropriate model is the 
Kratky-Porod chain \cite{kratky:49} which - apart from 
the radius of gyration 
\cite{saito:67} - can be handled only in certain 
limiting cases \cite{kroy:96,wilhelm:96},
leading 
to new difficulties, regarding
the conformational properties only. In order to keep 
the purely conformational part of the 
model as simple as possible, we confined ourselves to simple random walk
rings. The latter point is legal, since the topological properties discussed
above do not depend on the polymer chain model.

In this letter we presented a simple model for topological interactions in DNA
catenanes based upon the Gauss invariant as a minimal model for topology
conservation. The topological constraint of fixed linking number was expressed
in terms of an $n_{p}$-component Abelian Chern-Simons field theory and was 
applied to the case $n_{p}=2$. An alternative approach using Abelian BF-theory 
(see \cite{blau:91}) will be presented elsewhere. Using a pre-averaging
procedure and considering the effective topological interactions to lowest 
order in the variable $s_{\alpha\beta}$ (conjugate to the linking number 
$m_{\alpha\beta}$ between rings $\alpha$ and $\beta$), an effective free
energy 
was obtained for catenanes where the positions of one segment on each ring 
${\bf r}^{\alpha}_{0}$ and ${\bf r}^{\beta}_{0}$ were kept fixed. Treating the resulting
$R^{4}$ term which is both attractive and strongly anharmonic as the dominant
one, we obtained distribution functions for distances between segments located
on different rings. The 2D projections of these distribution functions show a
good qualitative agreement with recent experimental data on the conformation 
of DNA catenanes with different linking numbers \cite{levene:95}.

%

\newpage
\noindent
FIG. 1: The scaled elastic free energy $\beta
F_{\alpha\beta}(4/m^{2}_{\alpha\beta})$. \\

\noindent
FIG. 2: Distribution functions for the distance (in the projection plane) 
between
segments on separate rings for linking numbers $m=3, \;5,\; 7,\; 9$.

\noindent
FIG.3
The experimetal result drawn from electron micrographs. It is not sure that
these values correspond to thermal equilibrium. Thus a direct fit has been
avoided. 

\newpage
\begin{figure}[h]
  \begin{center}
  \begin{minipage}{15cm}
\epsfxsize 15cm
\epsfbox{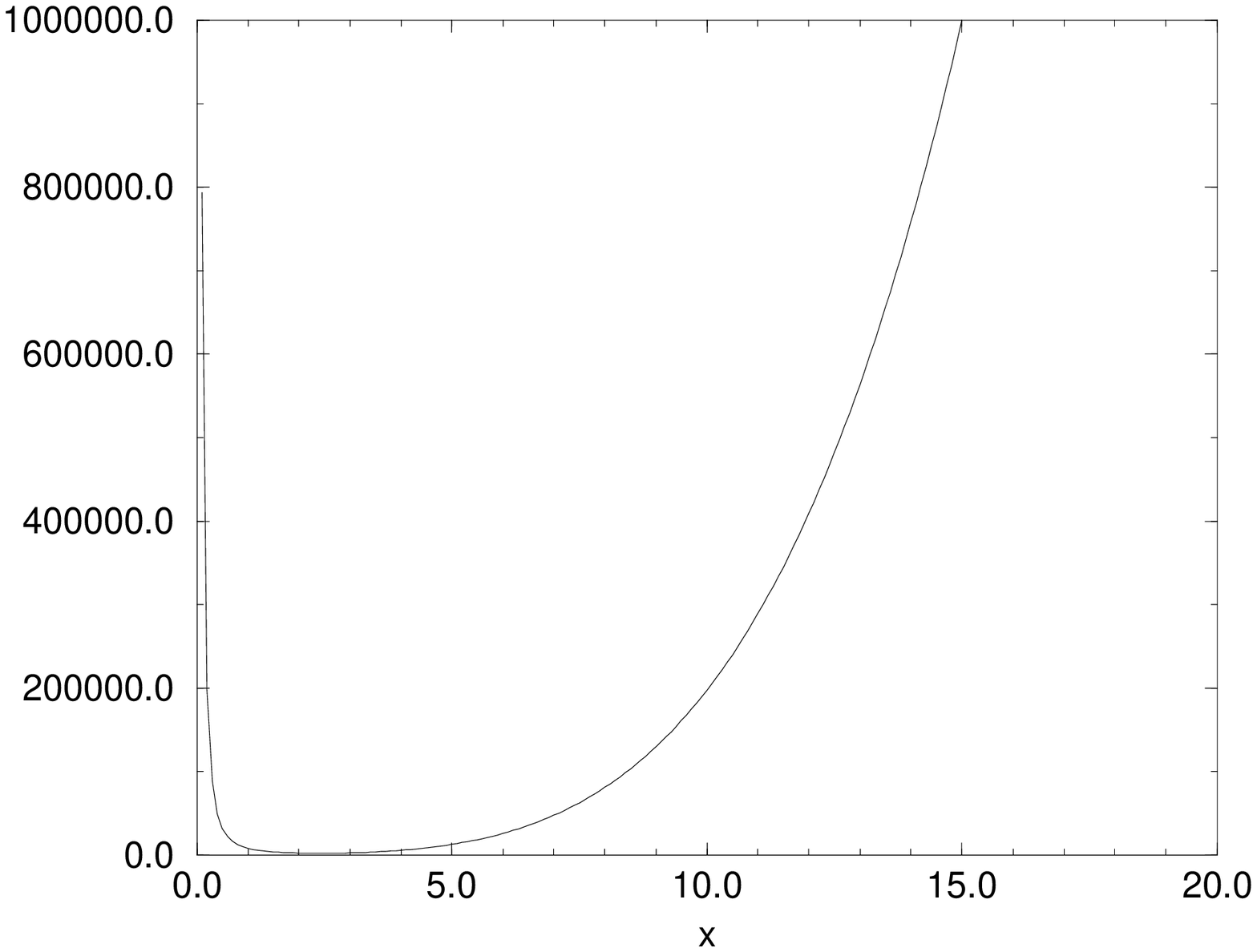}
  \end{minipage}
  \end{center}
  \label{ringpot}
\end{figure}
\newpage
\begin{figure}[h]
  \begin{center}
  \begin{minipage}{15cm}
\epsfxsize 15cm
\epsfbox{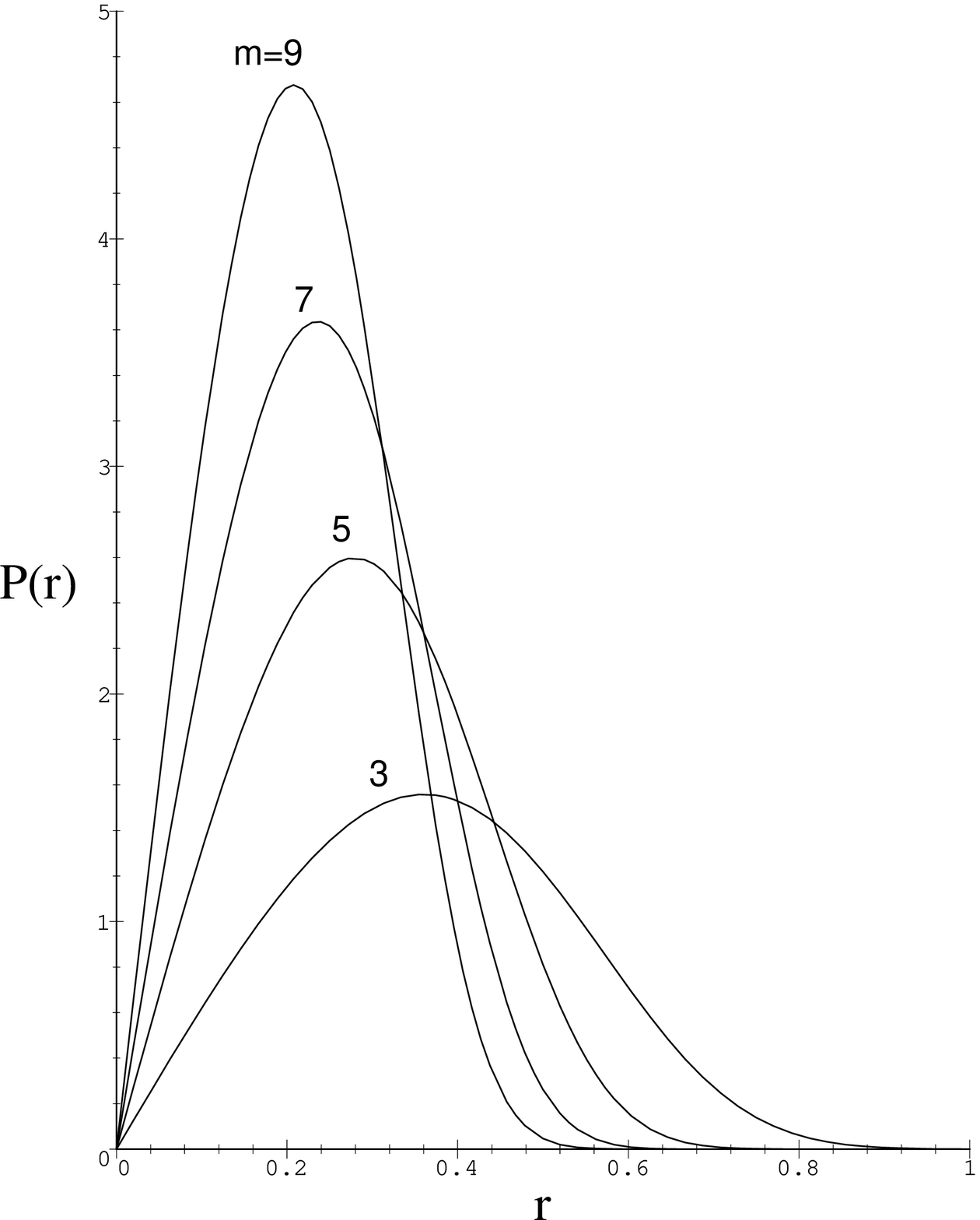}
  \end{minipage}
  \end{center}
  \label{dnavert}
\end{figure}

\newpage
\begin{figure}[h]
  \begin{center}
  \begin{minipage}{15cm}
\epsfxsize 15cm
\epsfbox{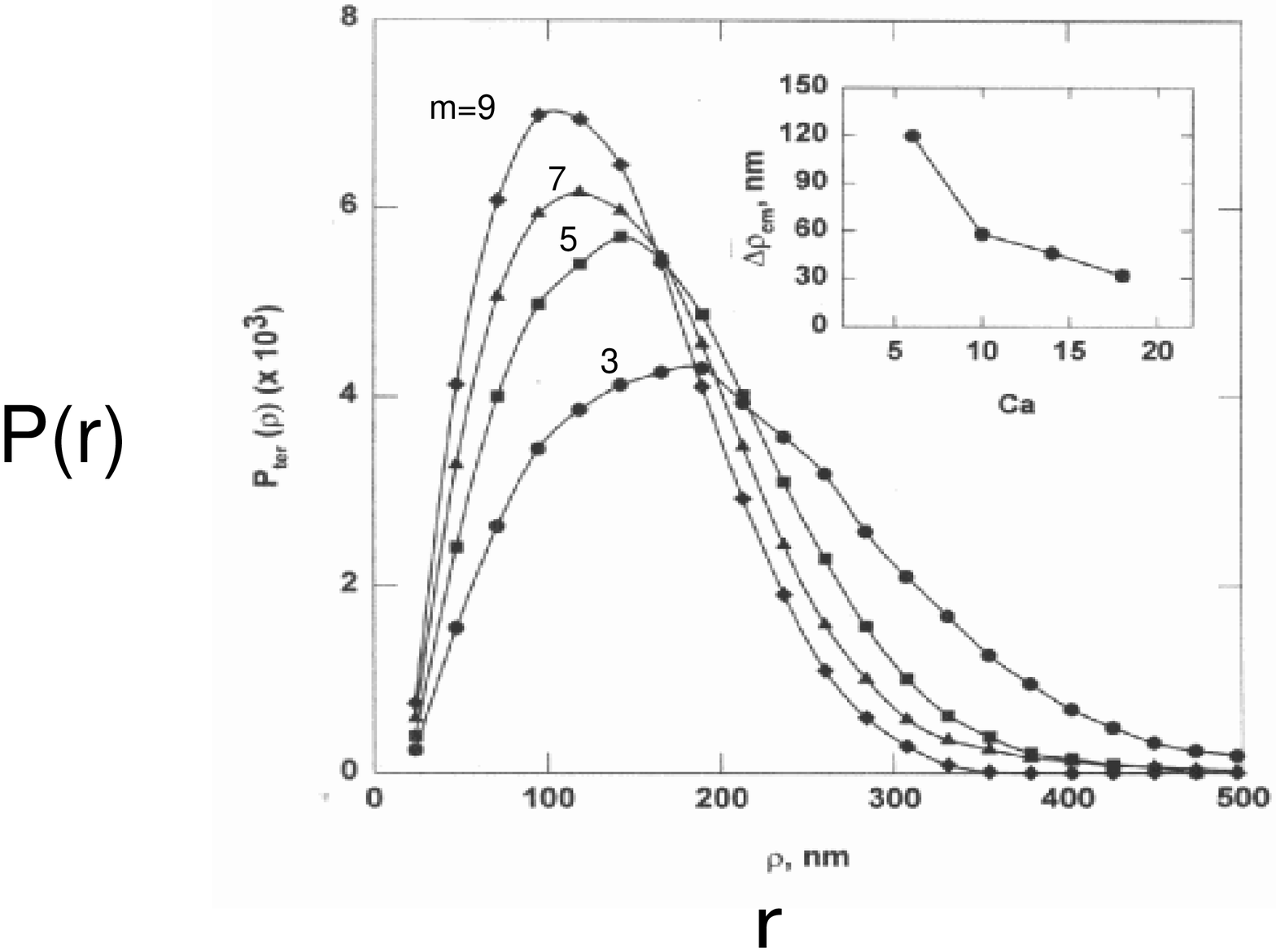}
  \end{minipage}
  \end{center}
  \label{ringpot}
\end{figure}

\end{document}